\documentstyle [12pt]{article}
\begin{document}

\begin{titlepage}
\begin{Large}
\begin{center}
{\bf Pion Form Factor: Transition From Soft To Hard QCD}
\end{center}
\end{Large}
\vspace{5 mm}

\begin{large}
\begin{center}
{L. S. Kisslinger and  S. W. Wang }
\end{center}
\end{large}
\vspace{5 mm}
\begin{center}
{Physics Department, Carnegie-Mellon University, Pittsburgh, PA 15213, USA }
\end{center}

\begin{large}
\begin{abstract}
We have reexamined the elastic pion form factor over a broad range of
momentum transfers with the mass evolution from current quark to
constituent quark being taken into account.  We have also studied the
effect of Sudakov form factors, of anomalous quark magnetic
moments and of alternative soft wave functions.
Our calculation shows a power-law falloff from the present
experimental values to near-asymptotic values in the few $GeV^{2}$ range.

\end{abstract}
\end{large}
\end{titlepage}

\section{Introduction And Review Of Method}

\hspace{3mm}
Quark and gluon structure functions of hadrons have been extracted via
inclusive processes, using experimental data at a few $GeV^{2}$ and above.
In the analysis of inclusive processes at $Q^{2}$
of a few $GeV^{2}$ and above QCD enters mainly in the evolution of the
parton distribution functions and in higher-twist scaling violations,
both of which are generally treated in perturbative QCD (PQCD).
In contrast to exclusive processes, both nonperturbative and perturbative
QCD are needed explicitly for the treatment of inclusive processes even
at very high $Q^{2}$.  It is important to use exclusive processes for the
study of the quark and gluon structure of hadrons, and to learn more
about the nature of QCD, but there are a number of theoretical problems
which must be solved.

It has been suggested\cite{th:pp} that
elastic form factors can be calculated
neglecting transverse momentum and
using a simple transition
operator arising from single-gluon exchange between quark pairs--the "hard
scattering" model--for $Q^{2}$ above a few $GeV^{2}$ (about 1 $GeV$ for
the pion), i.e., at momentum transfers where inclusive form factors scale.
The $Q^{2}$ dependence of the elastic form factors
predicted by the hard scattering model is consistent
with present experimental data for  $Q^{2}$ greater than about 1 $GeV^{2}$
for the pion and 3 $GeV^{2}$ for the nucleon, respectively.  However, the
magnitudes of both elastic form factors are in strong disagreement with the
known theoretical large-$Q^{2}$ limit by a factor of three for the pion and
about two orders of magnitude for the nucleon at the largest $Q^{2}$ for
which there is reliable data.  This raises serious questions about the
validity of the hard-scattering model for the $Q^{2}$ regions which have
now been reached in experiments.

Moreover, one essential assumption of the hard scattering model is that
``soft'' contributions are negligible.  This is a very controversial
subject\cite{is:pp}.  Since the soft contributions which
are calculated in quark models are determined mainly by the radius of the
pion, it is known that the soft contributions are consistent with experiment
at values of $Q^{2}$ at which the hard scattering model is often applied.
The hard scattering model is clearly inconsistent at that point.  There
is an extensive body of literature on the general subject of exclusive
processes and QCD, and we shall discuss a number of the most important
points for the case of the pion elastic form factor in the present work.

We have been developing a comprehensive picture based on the light-cone
Bethe-Salpeter (LCBS) approach\cite{ja:pp,kw:pp} for the treatment of the
elastic pion form factor.
The elastic pion form factor is related to pion electromagnetic current by
the following equation
\begin{equation}
 <P' \mid J^{\mu}(0) \mid P> = F_{\pi}(Q^{2})(P+P')^{\mu}
\label{eq:pform}
\end{equation}
In a quark-antiquark LCBS picture, this can be expressed as a convolution
integral in terms of LCBS amplitudes, $\Psi (x,{\bf k_{\bot}})$:

\label{sec-qform}
\begin{equation}
<P'\mid J^{+}(0)\mid P>  =\sum_{i,j} \int \frac{[dx][d^{2}k_{\bot}]}
{16 \pi^{3}}\Psi ^{\ast}(x,{\bf k'_{\bot}}) \Psi (x,{\bf k_{\bot}}) \frac{
\bar{u}_{i}}{\sqrt{k'^{+}_{i}}} \Gamma ^{+} \frac{u_{j}}
{\sqrt{k_{j}^{+}}}
\label{eq:qform}
\end{equation}
where $\Gamma ^{+} =f_{q1} \gamma ^{+} +  \frac{i}{2m_{q}}
\sigma ^{+\nu} t_{\nu} f_{q2}$ is the quark vertex function, and
$f_{q1}(t^{2})$, $f_{q2}(t^{2})$ are quark form factors.
These quark form factors are normalized
by $f_{q1}(t^{2}\!=\!0)=\!e_{q}$ (quark charge), $f_{q2}(t^{2}\!=\!0)=\!
\kappa _{q}$ (quark anomalous magnetic dipole moment).
We take  $\kappa _{q}$ as a parameter.
The LCBS equation has the form
\begin{equation}
\Psi(x,{\bf k_{\bot}}) = \int dx'd^{2}k'_{\bot}K(x,{\bf k_{\bot}};x',
{\bf k'_{\bot}})\Psi(x',{\bf k'_{\bot}})
\label{eq:lcbs}
\end{equation}
where $K$ is the kernel.  The hard scattering model can be obtained from Eqs.
(\ref{eq:qform},\ref{eq:lcbs}) by using the one gluon exchange operator
for $K$ and neglecting transverse momenta\cite{th:pp}.

For a comprehensive program starting from the LCBS approach
one needs a realistic kernel which includes both the confining and
asymptotic freedom aspects of QCD.  In our early work\cite{ja:pp}
we used the form

\begin{equation}
\Psi(x,{\bf k_{\bot}}) = \int dx'd^{2}k'_{\bot}[K_{conf}(x,{\bf k_{\bot}};
x',{\bf k'_{\bot}})
 + K_{ge}(x,{\bf k_{\bot}};x',{\bf k'_{\bot}})]\Psi(x',{\bf k'_{\bot}})
\label{eq:kw1}
\end{equation}
and used a relativistic string for the confining kernel, $K_{conf}$,
while the gluon exchange kernel, $K_{ge}$ is obtained from light-cone PQCD
\cite{th:pp}.  The confining kernel is not known, and even if an accurate
phenomenological form were determined through study of the pion, it could
not be used for the nucleon, which probably contains important three-
quark interactions.  In our recent work\cite{kw:pp} we have
avoided the difficult
problem of finding a phenomenological confining potential by using a
model for the soft amplitude.  I.e., we recognize that the soft BS
amplitude, $\Psi^{s}(x,{\bf k_{\bot}})$ can be considered to be the
solution of the equation

\begin{equation}
\Psi^{s}(x,{\bf k_{\bot}}) \equiv \int
dx'd^{2}k'_{\bot}K_{conf}(x,{\bf k_{\bot}};x',{\bf k'_{\bot}})
\Psi^{s}(x',{\bf k'_{\bot}})
\label{eq:kw2}
\end{equation}
Iterating Eq.(\ref{eq:kw1}) by inserting $\Psi^{s}$ for $\Psi$
one obtains the approximate form

\begin{equation}
\Psi(x,{\bf k_{\bot}}) \approx \Psi^{s}(x',{\bf k'_{\bot}}) +
\int dx'd^{2}k'_{\bot}
K_{ge}(x,{\bf k_{\bot}};x',{\bf k'_{\bot}})\Psi^{s}(x',{\bf k'_{\bot}})
\label{eq:kw3}
\end{equation}
By using this technique, one can obtain an
approximate solution of the Bethe-Salpeter (BS) equation, starting with a
relativistic bound state light-cone model wave function. This BS wave
function contains both soft and hard ingredients needed to take care of
momentum transfer for all $Q^{2}$ therefore is correctly characterized as
including both confinement and asymptotic features
of a composite quark system. Inserting the form of Eq. (\ref{eq:kw3}) in
Eq. (\ref{eq:qform}), one obtains the soft form factor (impulse approximation)
from the first term in Eq. (\ref{eq:kw3}), a generalization of the hard form
factor (the form of the hard form factor with transverse momentum effects
retained), and further correction terms.

The application of this approach to the pion
form factor has been shown\cite{kw:pp} to be in good agreement
with the direct BS calculation \cite{ja:pp}. We have predicted a power-law
falloff behavior for the form factor $Q^{2}F_{\pi}(Q^{2})$, which reaches
asymptotic limit at about 15 $GeV^{2}$.  A check on the next iteration of
Eq. (\ref{eq:kw1}) using the form of Eq. (\ref{eq:kw2}) has shown that the
correction is small when used in Eq. (\ref{eq:qform}) to calculate the pion
form factor.  One nice feature of this approach is that one determines
the soft part and the hard part separately, so that one can determine the
transition from soft to hard QCD within this LCBS approach.

In this letter, we are going to reexamine our model wave function in the
light of recent work by a number of other theorists.   In particular, we
carry out a study of 1) the evolution of quark mass with momentum,
2) the effect of the quark anomalous magnetic form factor,
3) the effect of the Sudakov form factor\cite{su:pp} and 4) the
question of the form of the wave function in the light of the proposed
form of Chernyak and Zhitnitsky\cite{cz:pp}, which has been used in
attempts to fit experiment at rather low $Q$ with the hard scattering
form, neglecting the soft contribution.

\section{Quark Mass Evolution}

\hspace{3mm}
The constituent quark model (CQM) has been successful in explaining many
static and low momentum transfer phenomena. But one of the deficiencies
of a CQM is that it totally ignores the dynamic aspects of quark
masses. We have known for a long time \cite{le:pp} that
at high energy (light) quarks are almost massless.  On the other hand,
the hard scattering model uses massless quark propagators or current quark
masses even when applied at rather low momentum transfer.

Since we are interested in getting information for all $Q^{2}$,
it is essential for us to understand the transition from the constituent
to the current quark mass. Analyses based on QCD sum rules
\cite{sh:pp,la:pp,po:pp} have indicated that quark mass evolution involves
both quark and gluon condensates.  The momentum dependence of the light
quark masses which results from these studies is of the form

\label{sec-mass}
\begin{equation}
m(p^{2})=[m+g^{2}( \frac{c_{1}}{p^{2}}+ \frac{(3mp^{2}+4m^{3})}{(p^{2}+
m^{2})^{3}}c_{2}]
\label{eq:mass}
\end{equation}
where $c_{1} \sim <\!\bar{\psi} \psi \!>$ (the quark condensate), while
$c_{2} \sim < \! G_{\mu \nu}^{2} \!>$ (the gluon condensate).

We can see from  Eq. (\ref{eq:mass}) that as its momentum increases, the quark
mass becomes smaller. At the limit where $p^{2}$ goes to infinity, $m(p^{2})\!
=\!m$ (current quark mass). Although Eq. (\ref{eq:mass}) is only valid above a
certain momentum scale, because the sum rule method used in
Refs.\cite{sh:pp,la:pp,po:pp} are not valid in very low momentum region,
we can still extract some useful information for
the mass evolution out of it. By a direct comparison and through calculations
we find almost identical numerical results for pion form factor
calculation for $Q^{2} > 2\, GeV^{2}$ by using either Eq. (\ref{eq:mass})
or the following form for the quark mass:

\label{sec-mass1}
\begin{equation}
m(p^{2})=m+(M-m) \frac{1+exp(-\frac{\mu }{\lambda ^{2}})}
{1+exp(\frac{-p^{2}-\mu}{\lambda ^{2}})}
-4m \frac{p^{2}}{(1-p^{2})^{2}}
\label{eq:mass1}
\end{equation}
where {\em m} and {\em M} are the current and constituent quark mass,
respectively. We take the parameters $\mu \! =\! 0.5 \: GeV^{2}$ and
$\lambda ^{2}\! =\! 0.2\:  GeV^{2}$. Eq. (\ref{eq:mass1}) is pure
phenomenological.  Nevertheless it gives
a satisfactory representation of the evolution of the quark mass from the
low energy limit of the constituent quark mass to the high energy limit
or the current quark mass.

Using Eq. (\ref{eq:mass1}) for the mass evolution, we calculate the
elastic pion form factor in a broad range of momentum transfer
($0 \le Q^{2} \le 60\: GeV^{2}$). The results are quite interesting.
They are shown in Figs. 1 and 2.
Firstly, the quark mass evolution causes only small differences
in the calculated results for the pion charge radius, $<\! r_{\pi}^{2}\! >$,
and the pion decay constant. Secondly,
The soft form factors at low momentum transfers ( $Q^{2} < 0.5\: GeV^{2}$)
are basically intact though there appears a tiny enhancement around
$Q^{2} \sim 0.5-2\: GeV^{2}$ region. However, the mass effect becomes
increasingly important for larger $Q^{2}$ in that the hard tail from soft
contribution is totally suppressed beyond $Q^{2} > 10\: GeV^{2}$. In other
words, the hard processes dominate completely above $\sim 10 \:GeV^{2}$.
Finally, the running quark mass seems to boost the hard contribution even
more at large $Q^{2}$, keeping in mind that the hard wave function is obtained
from Eq. (\ref{eq:kw3}) where quark mass is one of the essential parameters.

The momentum dependence of the quark mass is one of the most important
issues considered in the present work.  The general feature of a power
falloff from the present experimental values found in Ref.\cite{kw:pp}
is still found. The suppression of the soft contribution and the the
enhancement of the hard scattering at large $Q^{2}$ have brought
$Q^{2}F_{\pi}(Q^{2})$ very close to its asymptotic value as $Q^{2}$ goes
above 10 $GeV^{2}$. This is in good agreement with our previous work
although it probably takes a much higher $Q^{2}$ for one to reach the exact
asymptotic value.  We still conclude that there will be a power-law
falloff of $Q^{2}F_{\pi}(Q^{2})$ in the region between about 3 and 10
$GeV^2$.

\section{Quark Magnetic Form Factor}
\label{sec-pform}

\hspace{3mm}
Before we consider the effect of the momentum dependence of quark form
factors, we study the effect of a possible quark anomalous magnetic moments
by assuming that $f_{q1}\!=\!e_{q}$ and $f_{q2}\!=\! \kappa _{q}$.
It was suggested by Chung and Coester \cite{ch:pp} that quark anomalous
magnetic form factors would play important role in fitting the charge
radius and electromagnetic form factors of nucleon. It would be more
interesting to see if this is the case for the pion since pion wave function
can be more precisely determined for we have known the experimental value
of pion decay constant $f_{\pi}$ very well. The calculation only involves the
soft part, so that we use $\Psi^{s}$ in Eq. (\ref{eq:qform}).

The spin component of the wave function can be obtained by conducting a light
cone boost on a Melosh rotated quark-antiquark coupled state.
See\cite{ch1:pp}, for instance, for a detailed discussion.
The term involving quark magnetic moment's contribution to the form factor
can be explicitly written down as

\label{sec-form}
\begin{equation}
F_{\pi}(Q^{2}) \sim \int \frac{[dx][d^{2}k_{\bot}]}{16 \pi^{3}}(A-B)\frac{t}
{m}\kappa
\label{eq:form}
\end{equation}
where $\kappa = \kappa _{u} - \kappa _{\bar{d}}$, {\em m} is quark mass,
$t^{2}=(1-x_{1})^{2}Q^{2}$, {\em A} and {\em B} are
$\Psi ^{'\ast}_{\downarrow \downarrow}\Psi_{\uparrow \downarrow}+
\Psi ^{'\ast}_{\downarrow \uparrow}\Psi_{\uparrow \uparrow}$  and
$\Psi ^{'\ast}_{\uparrow \uparrow}\Psi_{\downarrow \uparrow}+
\Psi ^{'\ast}_{\uparrow \downarrow}\Psi_{\downarrow \downarrow}$,
respectively.

We have tried to adjust $\kappa$ to fit the experimental data at low
momentum transfers and found that it does improve the pion charge radius
$< \!r_{\pi} ^{2}\!>$ by about 5\% (with $\kappa \sim$ 0.01-0.05) given the
quark mass {\em m} and harmonic oscillator parameter $\alpha$. This seems to
indicate that at low $Q^{2}$, there indeed exists an effective quark anomalous
moment although the corrections are only within a few per cent as far as the
model wave function we choose is concerned. Nevertheless, the range of
$\kappa$
we determined here can serve as a guide for further theoretical and
experimental investigations of the quark anomalous moment. At large
$Q^{2}$ the quark anomalous moment is negligible.

\section{Sudakov Form Factor Effects}

\hspace{3mm}
There has been a great deal of interest in the introduction of Sudakov
form factors in the PQCD treatment of exclusive processes.  It has been
pointed out by many authors that the inclusion of such form factors is
necessary for consistency of the hard scattering assumption.  This
has been studied in detail by Sterman and coworkers, who give references
to earlier work.\cite{st:pp}

The Sudakov effect\cite{su:pp} results in the presence of double-logs in
vertex functions.  We introduce these effects by taking the form of the
quark form factor as\cite{bu:pp}

\begin{equation}
f_{q1}(q^{2}) =exp(-(C_{F}g^{2}/8 \pi^{2})ln \lambda ln \tau)
\label{eq:su}
\end{equation}
where $\lambda=k_{\bot}^{2}/k_{\bot}^{'2}$ and $\tau =Q^{2}/k_{\bot}^{'2}$.
We find that due to the running coupling constant in QCD the Sudakov
suppression is very mild in the domain of $Q^{2}$ where experimental data
are available. There is little change in our numerical results from
this effect.  This is consistent with recent calculations\cite{jak:pp}
which show that the inclusion of transverse momentum in the soft wave
function generates a suppression much stronger than the Sudakov to the hard
scattering in the region of present experiment.  These results are all very
similar to the observations in our previous work\cite{kw:pp}. We conclude
that it is more essential to take the transverse momentum in the soft wave
function into account.\footnote{ The quark distribution amplitude of ours is
very close to the asymptotic form. It is known that the Sudakov effect is
small for the asymptotic form of the quark distribution
amplitude.\cite{st:pp,jak:pp}}

\section{Form of Wave Function}

\hspace{3mm}
The evaluation of the hard scattering form for the form factor requires
a quark distribution amplitude.  As mentioned above, if one uses the
asymptotic value of the quark distribution amplitude or our model, the
hard scattering contribution is much smaller than the experimental values
of the pion form factor at the largest values of momentum transfer for
which there are measurements.
There are many published papers which suggest
that with a form such as that suggested by Chernyak and
Zhitnitsky\cite{cz:pp}, based on a QCD sum rule analysis,
one can use the hard scattering form at rather low $Q^{2}$.
Since the method of QCD sum rules does not make use of explicit
models of hadronic wave functions, this has been used to justify
the hard scattering model for the pion (and nucleon) elastic form
factors to fit present experiment.

We would like to make two observations about the quark distribution
amplitude (QDA). Our spin wave function, apart from an overall factor,
is identical to that proposed by Dziembowski\cite{dz:pp}. It has been
known that this type of wave function, when the parameters are specifically
chosen, can generate a corresponding quark QDA that is very similar to the
double-humped QDA suggested by Chernyak and Zhitnitsky based on
QCD sum rules analysis. One of our findings in this pion form factor
calculation is that, under the physical constraints imposed on the wave
function, one can never reach this special QDA. Our QDA is more like the
asymptotic one (see Fig. 3 for an illustration). This finding is consistent
with the recent analyses from both lattice calculation and QCD sum
rules\cite{dgr:pp}.

Secondly, the method of QCD sum rules extracts moments of distributions
rather than distributions, and these moments contain errors.  It has
recently been pointed out\cite{ehg:pp} that if one takes the errors into
consideration one cannot actually distinguish between the very asymmetric
quark distribution amplitude of Ref\cite{cz:pp} and a symmetric one such as
that resulting from our model, or even the asymptotic one.

We would like to add another comment here on the soft wave function we used
as the starting point to get our approximate solution for BS equation. This
concerns the difference of spin wave functions between the light cone
form and the instant form. For the pion, we first couple a valence quark and
an antiquark in the pion rest frame. A Melosh type Wigner rotation
transforms a state from its instant form to the light cone representation.
An important feature of this transformation is that it generates two extra
helicity components,  namely $h_{1} \pm h_{2}=\pm1$, apart from the
conventional helicity zero components. Contrary to the claims made by several
authors\cite{jak:pp,gg:pp} regarding the impossibility of properly fitting
the pion charge radius while still satisfying the constraints on the
pion wave
function imposed by the $\pi \rightarrow \mu \nu$ decay and $\pi^{0}
\rightarrow \gamma \gamma$ process, we would like to point it out that the
unconventional helicity components customarily neglected turn out to be
crucial in consistently fitting all the three constraints that the wave
function must satisfy. With the parameters given as: $m_{q}\!=\!0.33\: GeV,
\alpha \!=\!0.32\: GeV, \kappa = (\kappa_{u}-\kappa_{\bar{d}})=0.04$ and
$\Lambda_{QCD}=150\: MeV$, we get: $<\!r_{\pi}^{2}\!>=0.45 \:fm^{2}, f_{\pi}
=93.4\: MeV$. The analysis
of the contribution to the pure hard process from these unconventional
components is under way. We expect some interesting impact on the hard form
factors in the region where the PQCD is believed to be dominant. We will
report it later in a separate publication.

\section{Conclusion}

\hspace{3mm}
We have calculated pion elastic form factors at momentum transfer from
0 to 60 $GeV^{2}$ using the theoretical approach we developed in
\cite{kw:pp}. Four modifications of our model wave function have been made
and tested.
An important new result is that quark mass evolution is essential for
any theoretical models which expect to include both confinement and
asymptotic features in the theory.
An empirical formula for the quark mass produces an
identical effect for the pion form factor at $Q^{2} > 1\: GeV^{2}$ as QCD sum
rules do and is suitable for extensions to low momentum regions.
Our numerical calculation has shown that the effective  quark
anomalous magnetic moment could show up at low momentum transfer if such a
moment exists, but it has a very small effect on the pion form factor in
regions of interest for the transition from soft to hard QCD.
We have also studied the Sudakov form factor of quarks.  We
find that a Sudakov form factor has very little effect for the pion form
factor at the region of $Q^{2}$ where experimental data are available
currently.  We have also studied effects of the form of the soft wave
function.

While there is still no agreement in the value of the momentum transfer at
which one can reach the asymptotic limit of PQCD for exclusive processes,
our work seems to suggest that for the elastic scattering with a pion
target the hard contributions enter at about 1 $GeV^{2}$, become dominant
after about 10 $GeV^{2}$ and approach the asymptotic limit at $Q^{2}$ no
lower than 60 $GeV^{2}$. Experiments for momentum transfers in the range
between 4 to 15 $GeV^{2}$ will be a crucial test for our prediction,
since we predict a power-law falloff. More
accurate data at $Q^{2}= 1 \sim 4\: GeV^{2}$ are also needed for better model
building, for which CEBAF may play a significant role in the near future.

{\em Acknowledgement}: This work is supported in part by NSF grant
PHY-9023586.

\newpage

\begin{figure}
\caption{ Quark mass effect on form factors. The solid and dash
lines are those with or without quark mass evolution, respectively.
The experimental data are taken from\protect\cite{be:pp}.}
\end{figure}
\begin{figure}
\caption{ Pion form factors. The parameters are: $m_{q}\!=\!0.33\: GeV,
\alpha \!=\!0.32\: GeV, \kappa\! =\!0.04$ and
$\Lambda_{QCD}=150\: MeV$.
The experimental data are taken from\protect\cite{be:pp}.}
\end{figure}
\begin{figure}
\caption{ The quark distribution amplitude. The solid, dash-dot and dash
line curves correspond to the harmonic parameters 0.3, 0.55 and 0.58,
respectively. We take $m_{\pi} =\! 613\: MeV$\protect \cite{is:pp,dz:pp}
here merely for an illustration. (For actual calculation of the form factor
we use $m_{\pi}^{2}= (k_{\bot}^{2}+m^{2})/x(1-x)$, with which the
distribution function is about the same as the solid line shown above.)}
\end{figure}
\end{document}